\documentstyle[epsf,preprint,prb,aps]{revtex}
\def\a{\alpha}			    
\def\lb{\Lambda}
\def\d{\Delta}
\def\la{\langle}
\def\ra{\rangle}
\def\w{\omega}
\def\e{\epsilon}
\def\g{\Gamma}

\begin{document}
  \draft
\title{Coherent tunnelling through two quantum dots with Coulomb interaction}
\author{P. Pals and A.MacKinnon}
\address{The Blackett Laboratory, Imperial College,
London SW7 2BZ, United Kingdom}
\maketitle

\begin{abstract}
The coherent conductance and current is calculated 
through two quantum dots using the Hubbard model for a single level per 
spin. The occurrence of negative differential conductance is demonstrated.
The Ohmic conductance is calculated for dots with equally spaced levels. 
Transport is determined by matching energy levels, even when they do not 
occur at the charge degeneracy points. 
\end{abstract}
\pacs{34.80.P, 73.20.D}


\narrowtext
\noindent
The advance of lithographical techniques on a nanometre scale in recent years 
has made it 
possible to study systems that were inaccessible to experimentation before.
This has opened the way to produce and investigate structures where the 
carriers are 
confined to one or even zero dimensions. This has given rise to the 
discovery of a number of novel effects. It has been shown by Reed {\it et al.} 
that discrete states can clearly be discerned in quantum dots, structures 
which have been confined in all three dimensions \cite{Reed}. 
These technological advances have made it possible to study the interplay 
between charge quantisation effects (Coulomb blockade) and size quantisation
effects (discrete energy levels).

Whereas there has been a considerable amount of experimental and theoretical 
study on the conductive properties of single quantum dots \cite{Been,Lee}
and single and double
metallic dots \cite{Amman2,Ruzin,Kemerink}, so far relatively 
little attention has been paid to the case of two quantum dots in series.

The transport properties of a single dot are insensitive to incoherent
scattering, provided that the broadening of the levels is small 
\cite{Payne,Weil}. However, when two dots are connected in series, this no 
longer holds true. In this paper the coherent case will be considered where
the phase-breaking rate is small compared to the tunnelling rate.
Recently, the importance of coherence for transport through a single dots has 
been shown explicitly by direct measurement of the phase of the transmission
coefficient \cite{Yacoby}.

\section{Method}
{
\noindent
The system under investigation consists of an interacting region of 
one or more quantum dots connected between leads or electron reservoirs of
specified chemical potential. In general the reservoirs will have different 
chemical potentials, causing a current to flow through the dots. This makes
it inherently a non-equilibrium process. Therefore the system will be 
best described using non-equilibrium Green's functions, now commonly 
referred to as the Keldysh method \cite{Keldysh,Rammer}. This formalism is not
only valid in the linear response regime, but also for higher bias
voltages. A full description not only requires knowledge of the retarded 
and advanced Green's functions, as in equilibrium, but also of the 
`distribution' Green's function $G^<$ between the reservoir and dot sites.

\begin{eqnarray}
G^r_{n,m}(t,t') &=& -i\Theta(t-t')\la\{c_n(t),c^{\dagger}_m(t')\}
\ra \\
G^a_{n,m}(t,t') &=& i\Theta(t'-t)\la\{c_n(t),c^{\dagger}_m(t')\}
\ra  \\
G^<_{n,m}(t,t') &=& i \la c_m^{\dagger}(t') c_n(t)\ra
\end{eqnarray}
where $c^{\dagger}$ and $c$ are the usual creation and annihilation 
operators. 
In steady state the diagonal elements of ${\bf G}^r - {\bf G}^a$ are 
proportional to the local density of states, whereas 
${\bf G}^<$ plays the role of the density matrix.
They are related by the distribution matrix ${\bf F}$, defined by
${\bf G}^< = - {\bf F}({\bf G}^r-{\bf G}^a)$. In equilibrium ${\bf F}$ is 
simply a scalar and is identical to the Fermi-Dirac distribution function.

Having introduced the non-equilibrium Green's functions, the obvious
next step is to find an expression for the Landauer formula \cite{Landauer}
relating the current to the local properties of the system, such as the 
chemical potentials of the reservoirs, the density of states and the average 
occupation of the dots. It is assumed that the left and right electron 
reservoirs are large enough to have well-defined chemical potentials
$\mu_L$ and $\mu_R$. Any
interaction effects in the reservoirs can be neglected as a result of the 
screening by the free flow of charge carriers. 
Transport proceeds by electrons hopping between the reservoirs 
and the interacting region. The Landauer formula can be written as \cite{Meir} 
\begin{equation}
J = {ie \over 2 h} \int d\w {\rm Tr} \left[
(f_L(\w) {\bf \g}_L- f_R(\w) {\bf \g}_R)({\bf G}^r-{\bf G}^a) + 
    ({\bf \g}_L -{\bf \g}_R)  {\bf G}^< \right] \label{landau}
\end{equation}
where $({\bf \g}_{L/R})_{nm} = 2 \pi \sum_{\a \in L/R} \rho_{\a}(\w) 
V_{\a,n} (\w) V_{\a,m}^{\ast}(\w)$ are matrices coupling the interacting
region to the reservoirs. $V_{\a,n}$ are the hopping potentials between 
the reservoirs and the interacting region and $\rho_{\a}$ is the density 
of states in the reservoirs. 
It is worth noting that the Green's functions in this formula are to 
be calculated from the full Hamiltonian including the reservoirs, even 
though the reservoirs are already present in the couplings ${\bf \g}_{L/R}$ 
of the current equation.

For negligible bias voltages, knowledge of the retarded 
(and hence the advanced) Green's functions allows the distribution Green's 
functions to be calculated. The next step is to choose a particular Hamiltonian
$H$ and to calculate the retarded Green's functions explicitly.
This can be done using the 
equation of motion method which
consists of differentiating the Green's function with respect to
time, thus creating higher order Green's functions.
In Fourier space this amounts to the following 
iteration rule ($\hbar=1$):
\begin{equation}
\w \la \la \hat c;c^{\dagger}_{m} \ra 
\ra = \la \{\hat c, c^{\dagger}_{m}\}
\ra + \la \la [\hat c,H];c^{\dagger}_{m} 
\ra \ra  \label{iter}
\end{equation}
where $\hat c$ can be any combination of different creation and 
annihilation operators and 
$G^r_{n,m}(\w)=\la \la c_{n};c_{m}^{\dagger} \ra \ra$ is the
Fourier transform of the retarded Green's function.
The above equation can be applied successively to 
produce multiple particle Green's functions. For a system of two reservoirs and
$N$ dots with $n$ energy levels a multiple particle operator consists of a 
product of at most $4 (n N+2)-1$ 
single particle operators, i.e. the
creation and annihilation operators for both spin orientations in each dot 
level or reservoir. Therefore, it is in principle possible to obtain a closed 
set of equations. Clearly, it is a daunting task to calculate and solve all
possible simultaneous equations, so one usually resorts to making some 
approximations in order to close the set of coupled equations. 

\section{Single Dot}
{
\noindent
In order to determine
a realistic, yet simple, set of approximations, consider the simple case 
of a single quantum dot where only one relevant level for each spin direction 
will be taken into account. Any lower levels are assumed to 
be permanently occupied and higher energy levels are always empty. 
The two relevant states in the dot are labelled by the electron spin 
quantum numbers $\a$ or $\bar \a$.
A tight binding Hamiltonian will be used with a Hubbard term $U$ describing the 
Coulomb interaction between the spin up and spin down electrons. 
This is analogous to the Anderson model of localised impurity
states in metals \cite{Anderson}.
\begin{equation}
H_1 = \sum_{k\a \in L,R} \e_{k\a}c_{k\a}^{\dagger}c_{k\a} +
\sum_{\a}\!\e_{\a} c_{\a}^{\dagger}c_{\a} 
+ Un_{\a} n_{\bar \a} + \sum_{k\a \in L,R} (V_{k\a} c_{k\a}^{\dagger}
c_{\a} + H.c) \label{Hamil}
\end{equation}
where $\e_{\a}$ and $\e_{k\a}$ are the single particle energy levels
in the dot and reservoirs respectively. Transport between dot and reservoir
proceeds by means of the hopping potential $V_{k\a}$. 
Any interaction that might take place with electrons in 
the full lower energy levels is incorporated in the definition of 
$\e_{\a}$. The hopping potentials $V_{k\a}$ are taken to be real, 
because of the time-reversal symmetry.
It is required that any set of approximations yields the correct solution 
in the limit of zero bandwidth, i.e. negligible hopping potential. Moreover,
the single particle density of states integrated over all energies must 
equal unity. The density of states must be non-negative at all energies. 
Finally, the solution should display electron-hole symmetry. It should be 
invariant under the transformation $\{n \rightarrow 1-n, U \rightarrow -U,
\e_{\a} \rightarrow \e_{\a}+U,f \rightarrow 1-f \}$. 

The crudest set of 
approximations (scheme 1) consistent with these conditions 
\cite{Hubbard} corresponds to the
Hartree approximation scheme \cite{Lacroix} where electrons of 
opposite spin do not interact directly, but only through a modified 
average potential field.  
The resulting Green's function is
\begin{equation}
\la \la c_{\a};c_{\a}^{\dagger} \ra \ra  = {\w - \e_{\a} 
- U(1 - \la n_{\bar \a} \ra) \over (\w - \e_{\a})(\w - 
\e_{\a} - U) - \Sigma_{0\a}[\w - \e_{\a} - U(1 - \la 
n_{\bar \a} \ra)]} \label{crude}
\end{equation}
with
\begin{equation}
\Sigma_{0\a}= \sum_{k \in L,R} {|V_{k\a}|^2\over \w - \e_{k\a}} 
= -i  (\g_L+\g_R)/2
\label{sigma0a}
\end{equation}
where $\g_{L,R} = 2 \pi \rho_{L,R} |V_{L,R}|^2$. 
$\Sigma_{0\a}$ is the self-energy due to the tunnelling of the $\a$ electron
and $\la n_{\bar \a} \ra$ is the average occupancy of the opposite 
spin state. The same Green's function results from
the method as indicated by Caroli, Combescot and Nozieres \cite{Caroli},
where the full Green's function is calculated as a perturbation of the
system without transmission to either reservoir.
The occupation is the
integrated product of the (diagonal) distribution function and the local 
density of 
states in the dot. Since the Green's function is related to the density
of states which in turn depends on the average occupation number,
it means that equation 
\ref{crude} is to be solved self-consistently. It can be shown that the 
integrated density of states equals unity, subject to the condition 
$0 \leq \la n \ra \leq 1$, which is obviously satisfied in any
real system. 

A better approximation scheme (scheme 2) is one which neglects the simultaneous
tunnelling of the electron of opposite spin, but which does not decouple the 
higher order Green's functions as in approximation scheme 1. 
This yields the following Green's function:
\begin{equation}
\la \la c_{\a};c_{\a}^{\dagger} \ra \ra  = 
{1-\la n_{\bar \a} \ra \over \w - \e_{\a} - \Sigma_{0\a}} +
{\la n_{\bar \a} \ra \over \w - \e_{\a} -U - \Sigma_{0\a}} 
 \label{scheme2}
\end{equation}
As the tunnelling of the opposite spin electron is not taken into account,
its creation and annihilation operators are always paired together in the 
higher order Green's functions. The operators for both spins do not 
truly intermingle, which results in the probabilistic expression for 
the Green's function. Therefore, the same expression can be obtained by 
solving the Green's function for the isolated site and adding an 
imaginary part to the site energy to include the reservoirs, i.e. the 
reservoirs simply serve as a source of broadening.

A more accurate method (scheme 3) only neglects terms involving correlation in 
  the leads, valid for temperatures higher than the Kondo temperature
\cite{Lacroix}.
The Green's function corresponding to these approximations is \cite{Lee}
\begin{eqnarray}
\la \la c_{\a};c_{\a}^{\dagger}\ra \ra &=& {1 -\la 
n_{\bar \a}
\ra \over\w - \e_{\a} - \Sigma_{0\a} + U\Sigma_{1\a}(\w - 
\e_{\a} - U - \Sigma_{0\a} - \Sigma_{3\a})^{-1}} \nonumber \\
&+&{\la n_{\bar \a} \ra \over\w - \e_{\a} - U - \Sigma_{0\a} 
- U\Sigma_{2\a}(\w - \e_{\a} - \Sigma_{0\a} -\Sigma_{3\a})^{-1}}
{\rm \qquad}
\end{eqnarray}
The self-energies due to the tunnelling of the $\bar \a$ electron
are defined as
\begin{equation}
\Sigma_{i\a}(\w) = \sum_{k \in L,R} A^{k \bar \a}_i |V^{k \bar \a}|^2  
\left({1\over\w - \e_{\a} + \e_{\bar \a} - 
\e_{k \bar \a}} + {1\over\w - \e_{\a} - \e_{\bar \a} - U + 
\e_{k \bar \a}}\right)  \label{sigmaia}
\end{equation}
where $A^{k \bar \a}_1 = f(\e_{k\bar \a})$, $A^{k \bar \a}_2 = 1 - 
f(\e_{k\bar \a})$ and $A^{k \bar \a}_3 = 1$.
As for scheme 2, it can be shown that the density of states integrates to 
unity for all values of $\la n \ra$.

Since the couplings to the reservoirs differ by a constant factor
the distribution matrix reduces to a diagonal matrix.
\begin{equation}
{\bf F} = {f_L \g_L + f_R \g_R \over \g_L + \g_R} \pmatrix{1 & 0 \cr 0 & 1\cr}
\end{equation}
As there is only one level for each spin 
the current expression \ref{landau} takes an especially simple form 
where the current is proportional
to the density of states between the Fermi levels of the reservoirs. 
\begin{equation}
J = {e \over \hbar} {\g_L \g_R \over \g_L + \g_R} \int d\w \rho(\w) 
\left[f_L(\w)-f_R(\w)\right]
\label{Jsimp2}
\end{equation}
The Ohmic conductance is given by the integrated product of the 
density of states and the derivative of the Fermi function.
At low temperatures the derivative of the Fermi function approximates 
a $\delta$-function. A Lorentzian density of states with its broadening 
caused solely by tunnelling to and from the reservoirs thus gives rise to a 
conductance peak of height $e^2/h$, provided that the dot is symmetrically 
coupled to the reservoirs.

The three approximation schemes are compared in figures \ref{fig:currsch}-
\ref{fig:occsch} for the case where the temperature equals the broadening due
to the reservoirs.  In
order to distinguish between the electrons in the dot, a magnetic field is
applied giving rise to a Zeeman splitting of the energy levels ($\e_{\a}=0.0 U,
\e_{\bar \a} = 0.1 U$). 

Figure \ref{fig:currsch} shows the current and the associated differential
conductance through the dot when a bias voltage is applied across it, keeping
the chemical potential in the right reservoir fixed  
($\mu_R = -0.5 U$). All three approximation schemes yield qualitatively the 
same results, although scheme 3 (and scheme 2 to a lesser extent) has a 
larger associated broadening. This is a direct consequence of the fact that 
scheme 3 takes the tunnelling of the oppositely charged electron into account.

In figure \ref{fig:condsch} the Ohmic conductance is plotted. 
The approximation schemes give
identical solutions for the shape and height of the major conductance peaks,
but especially approximation scheme 1 gives a different result for the 
amplitude of the minor peaks.
Moreover, an investigation of the occupation plot \ref{fig:occsch} shows that 
for scheme 1 the average occupation of the higher energy spin state can exceed 
that of the other spin state for intermediate values of the chemical potential
of the reservoirs. This is clearly unphysical. 
Both these observations justify the conclusion that scheme 1 should be 
dismissed as a good set of approximations.

As expected the conductance plot shows peaks at the single particle energy
levels and at the same energies displaced by $U$. It is clear that two peaks
are suppressed, which can be explained as follows. The first peak occurs 
when transport takes place through the lowest level $\e_{\a}$. 
As the chemical potential in the reservoirs is raised, the state $\e_{\a}$ 
will start to fill up. When the chemical potential 
lines up with 
$\e_{\bar\a}$, level $\e_{\a}$ will be mostly occupied (see figure 
\ref{fig:occsch}) thus putting the Coulomb blockade in place and impeding
the flow of electrons through level $\e_{\bar\a}$. This 
results in a highly suppressed conductance peak. Because of the 
electron-hole symmetry a similar situation occurs at much higher 
values of the chemical potential. 

Approximation schemes 2 and 3 seem to give results which are closely in 
accordance with each other. The main difference is that approximation scheme 3
produces a larger associated broadening. Although slightly better, scheme 3 is 
harder to manipulate numerically, whereas scheme 2 is much more stable.  
Therefore approximation scheme 2 will be employed here. 
}

\section{Two dots}
{
\noindent
The single dot Hamiltonian  can easily be extended 
to describe two quantum dots in series. As before, assume that there is
only one spin-split level per dot and that the electrons experience an
on-site interaction energy $U_{1/2}$.
\begin{eqnarray}
H_2 &=& \sum_{k\a \in L,R} \e_{k\a}c_{k\a}^{\dagger}c_{k\a} +
\sum_{\a} (\e_{1\a} c_{1\a}^{\dagger}c_{1\a}+\e_{2\a} c_{2\a}^{\dagger}c_{2\a}) 
\nonumber \\
&&+ U_1 n_{1\a} n_{\bar 1\a} + U_2 n_{2\a} n_{\bar 2\a} 
+ \sum_{\a} V_M (c_{1\a}^{\dagger} c_{2\a} + H.c.) \nonumber \\
&&+\sum_{k\a \in L} (V_{k\a} c_{k\a}^{\dagger} c_{1\a} + H.c.)
+\sum_{k\a \in R} (V_{k\a} c_{k\a}^{\dagger} c_{2\a} + H.c.)
\end{eqnarray}
As before the dots are coupled to the reservoirs by the hopping potentials
$V_{k\a}$, but there is also some coupling $V_M$ between the dots. 
In this representation the coupling matrices for each spin are given by
\begin{equation}
{\bf \g}_L = \pmatrix{\g_L & 0 \cr 0 & 0 \cr} {\rm \quad ; \quad}
{\bf \g}_R = \pmatrix{0 & 0 \cr 0 & \g_R \cr} \label{gsite}
\end{equation}
Now that there is more than one dot in the interacting region, it is clear
that the correlations between the occupation numbers of the dots become
important, since the occupation of one dot will depend on the occupation 
of the other. 
The equivalent of approximation scheme 1 actually ignores this correlation.
Here the approximation scheme 2 will be employed, where the reservoirs are 
included in the form of a self-energy. 
The single particle Green's function matrix for spin $\a$ is given by
\begin{eqnarray}
{\bf G}^r =     {\la    n_{1\bar\a}    n_{2\bar\a}  \ra  \over
            (\w-\e'_{1\a}-U_1)(\w-\e'_{2\a}-U_2)-V_M^2}
  &\pmatrix{\w-\e'_{2\a}-U_2 & V_M \cr V_M & \w-\e'_{1\a}-U_1 \cr}& 
\nonumber \\* 
  + {\rm \quad} {\la    n_{1\bar\a} (1-n_{2\bar\a}) \ra  \over
            (\w-\e'_{1\a}-U_1)(\w-\e'_{2\a}    )-V_M^2}
  &\pmatrix{\w-\e'_{2\a}     & V_M \cr V_M & \w-\e'_{1\a}-U_1 \cr}{\rm \qquad}& 
\nonumber \\*
  + {\rm \quad} {\la    n_{2\bar\a} (1-n_{1\bar\a}) \ra  \over
            (\w-\e'_{1\a}    )(\w-\e'_{2\a}-U_2)-V_M^2}
  &\pmatrix{\w-\e'_{2\a}-U_2 & V_M \cr V_M & \w-\e'_{1\a}     \cr}{\rm \qquad}& 
\nonumber \\*
  + {\rm \quad} {\la (1-n_{1\bar\a})(1-n_{2\bar\a}) \ra  \over
            (\w-\e'_{1\a}    )(\w-\e'_{2\a}     )-V_M^2}
  &\pmatrix{\w-\e'_{2\a}     & V_M \cr V_M & \w-\e'_{1\a}     \cr}{\rm \qquad 
\qquad}& 
\end{eqnarray}
where $\e'_{1\a} = \e_{1\a}-{i\over 2}\g_L,\e'_{2\a}=\e_{2\a}-{i\over 2}\g_R$.
As for the single dot case,
the Green's functions have a straightforward probabilistic interpretation. 
They treat the electrons of opposite spin $\bar \a$ as static entities 
whose main influence is to change the effective site energy of the 
electrons of spin $\a$ by $U$. Each Green's function is 
simply a sum over effective {\it non-interacting} Green's functions 
each weighted by the probability of a particular realisation of the 
occupation numbers for the opposite spin states. Of course the single 
particle energy levels have to be 
adjusted to correspond to the correct occupations of the opposite spin 
states.

This form of the Green's function has some important consequences.

\noindent $\bullet$ First of all the distribution, retarded and advanced 
self-energies can be written in terms of the coupling matrices: 
${\bf \Sigma}^< = i(f_L {\bf \g}_L + f_R {\bf \g}_R)$ and
${\bf \Sigma}^r-{\bf \Sigma}^a = -i({\bf \g}_L+{\bf \g}_R)$.
Substitution of the equalities ${\bf G}^< = {\bf G}^r {\bf \Sigma}^< {\bf G}^a$  
and ${\bf G}^r-{\bf G}^a = {\bf G}^r ({\bf \Sigma}^r-{\bf \Sigma}^a){\bf G}^a$
into equation \ref{landau} yields a condensed form of the current formula
for non-interacting systems.
\begin{equation}
J = {e \over h} \int d\w [f_L(\w)-f_R(\w) ] {\rm Tr} \left[
{\bf \g}_L {\bf G}^a {\bf \g}_R {\bf G}^r \right] \label{Jtrace}
\end{equation}
Using equations \ref{gsite} for the double dot this reduces to 
\begin{equation}
J = {e \over h} \sum_{\a} \int d\w [f_L(\w)-f_R(\w) ]  \g_L \g_R 
G^a_{12} G^r_{21}
\label{J2site}
\end{equation}

\noindent $\bullet$ Secondly, the coupling matrices 
are no longer proportional, which results in a non-diagonal form of the 
distribution matrix at finite bias. 
\begin{equation}
{\bf F} = {\bf G}^r \pmatrix{f_L & 0 \cr 0 & f_R \cr} ({\bf G}^r)^{-1}
\label{defF1}
\end{equation}

\noindent $\bullet$ Thirdly, the probabilistic form of the Green's functions 
warrants a statistical approach for the multiplication of Green's functions. 
Since the Green's functions are probabilistic sums over non-interacting
Green's functions, products of Green's functions must be expressed as 
probabilistic sums over the products of the appropriate non-interacting
Green's functions.

\noindent $\bullet$ Fourthly, higher order Green's functions, needed to 
calculate  correlations of the type $\la n_{1\a} n_{2\a} \ra$, can also be 
reduced to a sum over non-interacting Green's functions. 
For a non-interacting system it can be shown that
\begin{equation}
 \la n_1 n_2 \ra = \la n_1 \ra \la n_2 \ra - \la c_1^{\dagger} c_2 \ra
 \la c_2^{\dagger} c_1 \ra
\label{avocc}
\end{equation}
In the presence of interaction on the dots, the correlation $\la n_{1\a} 
n_{2\a} \ra$ is obtained by taking the probabilistic sum over all 
possible occupation realisations of the opposite spin states. 

Taking the four points above into account, the current and conductance 
characteristics can be calculated.  
Figure \ref{fig:cond2s} shows the Ohmic conductance through two dots 
which are identical in all respects apart from a relative 
energy off-set of $0.4 U$. In order to be able to interpret the 
conductance plots more easily the average occupation numbers for the 
four single-particle states are also plotted. 

In the limit of negligible coupling between the dots, the occupation for 
each dot is determined solely by the coupling to the adjoining reservoir. 
This is identical to the single dot case ({\it cf} figure \ref{fig:occsch}). 
The Ohmic conductance displays peaks at all the single particle energy levels
and at the same energies displaced by $U$. Similar to the conductance through
a single dot (see figure \ref{fig:condsch}), the peaks come in pairs 
separated by the Zeeman energy $\e_{\bar\a}-\e_{\a}$. As it has been explained 
for the single dot, one of the peaks of each pair is suppressed as a 
result of the Coulomb blockade in the dot. However, it is apparent from 
figure \ref{fig:cond2s}a that there is also a modulation of the height of 
pairs of peaks. For instance the first major peak is noticeably 
larger than the second peak. The current path that gives rise to the 
first major peak encompasses the levels $\e_{1\a}$ and $\e_{2\a}$. At 
the chemical potential at which the second major peak occurs, the first
dot will be virtually completely occupied and the Coulomb blockade will 
be in place. Therefore the dominant contribution to the current will 
come from the energy levels $\e_{1\a}+U$ and $\e_{2\a}$. In the case of 
figure \ref{fig:cond2s}a the dominant levels for the second peak are 
energetically further apart than those for the first peak, which explains
why the second pair of peaks is slightly suppressed. 
In short, the form of the conductance plot for negligible inter-dot coupling
$V_M$ can be understood by realising that electrons contribute to the 
transport only when the Coulomb barrier is overcome in both the first and 
the second dot.

As the coupling between the dots is increased, the energy levels in the dots
start to interact, repelling one another. 
The conductance peaks are shifted in energy. 
When the coupling between the dots becomes appreciable compared to the 
energy level difference $\e_{2\a}-\e_{1\a}$ for the decoupled dots, 
the number of peaks in the conductance can increase to a number that 
is greater than the total number of single particle energy levels. 
This is a result of the fact that the amount by which a level is repelled
depends on the energy difference with the level by which it is repelled.
For instance, an original level at an energy $\e_{1\a}$ can be repelled to 
two different resultant energies depending on whether it interacts with 
level $\e_{2\a}$ or $\e_{2\a}+U$. This causes a conductance peak both to 
shift in energy and to split. This is particularly significant when a 
level is repelled into opposite directions by the two possible levels it 
could interact with. This happens at intermediate values of the chemical 
potential. This explains why splitting can start to be observed for the 
middle peak pairs of figure \ref{fig:cond2s}c but not for the peaks at more
extreme values of the chemical potential. 

In figure \ref{fig:curr2s} the current through the two dots is plotted 
as a function of the chemical potential in the left reservoir, keeping the 
chemical potential in the right reservoir fixed. The I-V characteristics 
are studied as the coupling between the dots is increased. 

In the limit of negligible coupling the occupation of each dot is 
completely determined by the tunnelling of electrons to and from the 
adjoining reservoir. The barrier between the dots is clearly the current 
limiting segment, so that the current is proportional to $V_M^2$ (see 
figure \ref{fig:curr2s}a). Transport can proceed because the wavefunctions 
corresponding to the levels in the first dot can leak slightly into the 
second dot and vice versa, thus creating current paths. 
This explains why there is a marked increase in the current
at $\mu_L = \e_{1\a}$, $\mu_L = \e_{2\a}$ and 
$\mu_L = \e_{2\bar\a}$. Note that no current step occurs at 
$\mu_L = \e_{1\bar\a}$, even though an extra current path becomes 
available. This is most easily comprehended by drawing a distinction 
between the number of current paths and the number of effective 
current paths. A current path consists of a composite state which 
is a mixture of a state in the first dot and one in the second dot. 
A current path can be said to be effective when it can be used constantly. 
For example, it is true that a second current path becomes available 
at $\mu_L = \e_{1\bar\a}$. However, the number of effective paths 
remains fixed at $1$, since the Coulomb blockade in the first dot prevents 
the paths from being used simultaneously. 

The most notable feature of figure \ref{fig:curr2s}a is the region of 
negative differential conductance around $\mu_L = \e_{1\a}+U$. This 
is the energy at which the single particle levels in the first dot 
are both likely to be occupied. The dominant current 
contribution arises from the interaction of the levels $\e_1 + U$
with the levels $\e_2$. As these states are energetically further separated 
than the levels $\e_1$ and $\e_2$, this means that there is less overlap
of the wavefunctions in the two dots which accounts for the drop in current.

Finally, the more realistic case of the Ohmic conductance through a double 
dot with an infinite number of energy levels will be calculated. The simple 
case of negligible broadening and small inter-dot coupling will be considered.
The coupling to the reservoirs in the 'site' representation is given by 
the following matrices:
\begin{equation}
{\bf \g}_L = \left(\begin{array}{ccc} 
             \hat \g_L & \vdots & \hat 0 \\
             \stackrel{\textstyle \cdots}{}  & 
  \stackrel{\textstyle \cdot}{} \vdots \stackrel{\textstyle \cdot}{} 
                      & \stackrel{\textstyle \cdots}{}  \\
             \hat 0  & \vdots & \hat 0
             \end{array} \right)
{\rm \quad ; \quad}
{\bf \g}_R = \left(\begin{array}{ccc} 
             \hat 0 & \vdots & \hat 0 \\
             \stackrel{\textstyle \cdots}{}  & 
  \stackrel{\textstyle \cdot}{} \vdots \stackrel{\textstyle \cdot}{} 
                      & \stackrel{\textstyle \cdots}{}  \\
             \hat 0  & \vdots & \hat \g_R 
             \end{array} \right)
\end{equation}
The matrix is subdivided into four infinite sub-matrices, corresponding to 
coupling either between levels of the same dot or between levels of different
dots, e.g. the top left sub-matrix includes all couplings between levels of
the first dot. The sub-matrices $\hat \g_{L/R}$ are defined as a matrices whose 
elements are all given by $\g_{L/R}$. 
As before the current will be determined using equation \ref{Jtrace}. 
Substitution of the coupling matrices and taking the limit of an infinitesimal
(Ohmic) bias across the system yields the conductance equation.
\begin{equation}
G = {e^2 \over h} \int d\w {\g_L \g_R \over 
 4 k_BT \cosh^2 ({\w-\mu \over 2 k_BT})} \left|\sum_{i \in {\rm dot 1}} 
\sum_{j \in {\rm dot2}}
G^r_{ij}(\w) \right|^2
\end{equation}

\noindent
Analogous to expression \ref{J2site} the conductance only depends on the 
Green's functions between states in different dots. 
Treating the inter-dot coupling $V_M$ as a small perturbation allows these 
off-diagonal Green's functions to be written as 
$G^r_{ij}(\w) = G^r_i(\w) V_M G^r_j(\w)$ so that
\begin{equation}
G = {e^2 \over h} \int d\w {\g_L \g_R V_M^2 \over 
 4 k_BT \cosh^2 ({\w-\mu \over 2 k_BT})} 
\left|\sum_{i \in {\rm dot 1}} G^r_i(\w)\right|^2
\left|\sum_{j \in {\rm dot 2}} G^r_j(\w)\right|^2
\label{Gcohinf}
\end{equation}

\noindent 
In this approximation the diagonal elements $G^r_i(\w)$ of the Green's function
are identical to those of the single dot with multiple levels \cite{Lee}. Hence
\begin{equation}
\g_L \left|\sum_{i \in {\rm dot 1}} G^r_i(\w)\right|^2 = 
\g_L \left|\sum_{i \in {\rm dot 1}} \sum_{N_1} 
{P_i(N_1) \over \w-\e_i+i\g_L/2-N_1 U_1} \right|^2
\end{equation}
where $P_i(N_1)$ is the probability that the first dot contains $N_1$ electrons 
on levels other than $\e_i$. In the limit of negligible broadening this 
becomes
\begin{eqnarray}
\g_L \left|\sum_{i \in {\rm dot 1}} G^r_i(\w)\right|^2 &=&
2 \pi \sum_{i \in {\rm dot1}} \sum_{N_1} P_i(N_1) \delta(\w-\e_i-N_1 U_1) 
\nonumber \\
&+&  \g_L \left( \sum_{i \in {\rm dot1}} \sum_{N_1} 
{P_i(N_1) \over \w-\e_i-N_1 U_1} \right)^2
\end{eqnarray}
 
\noindent 
The first part of the expression shows the behaviour at energies coinciding
with an energy level whereas the second part gives the off-resonance 
contribution. A similar expression may be obtained for the Green's functions 
of the second dot. Note that the limit of small broadening excludes the 
possibility of
energy levels in the two dots matching up exactly (this case will be 
examined later in more detail), so that the conductance formula \ref{Gcohinf} 
can be rewritten as
\begin{eqnarray}
G &=& {e^2 \over h} \int d\w 
  {2\pi V_M^2 \over 4 k_BT \cosh^2 ({\w-\mu \over 2 k_BT})}  \nonumber \\
&\times&
\left[
    \sum_{i \in {\rm dot 1}} \sum_{N_1} \g_R P_i(N_1) \delta(\w-\e_i-N_1 U_1) 
\left(
    \sum_{j \in {\rm dot 2}} \sum_{N_2} {P_j(N_2) \over \w-\e_j-N_2 U_2}
\right)^2
\right.
  \nonumber \\
&+& \left.
    \sum_{j \in {\rm dot 2}} \sum_{N_2} \g_L P_j(N_2) \delta(\w-\e_j-N_2 U_2) 
\left(
    \sum_{i \in {\rm dot 1}} \sum_{N_1} {P_i(N_1) \over \w-\e_i-N_1 U_1}
\right)^2
\right] \qquad
\label{cohinfexp}
\end{eqnarray}

\noindent
In figure \ref{fig:cohinf} the Ohmic conductance has been plotted for 
a system where the dots have a different level spacing and Coulomb interaction.
As expected at low temperatures,  
the conductance peaks are largest at values of the chemical potential where the 
charge degeneracy points of the two dots are close in energy. At the charge 
degeneracy point between the occupations $N$ and $N+1$ the $(N+1)$th single
particle level is dominant.
As the temperature is raised there is a higher probability of other levels 
contributing to the current. This has 
a particularly strong effect when any of the newly available subsidiary levels
very nearly match up (as at $\mu \simeq -4.7U$). Otherwise a rise in 
temperature simply causes the peaks to be smeared. At higher temperatures 
the periodicity of the smaller dot $U_2+\d_{\e_2}$ can easily be estimated
from the spacing of the peaks in the conductance plot. 

At this stage it is useful to analyse the height and shape of the 
conductance peaks in more detail. Moreover, in order to compare the coherent
case with other tunnelling mechanisms, it would be helpful to obtain an 
expression for the current through the double dot at finite bias as a function
of the gate potential of one of the dots. 
At low temperatures these quantities can be investigated by just 
considering the dominant levels in each of the dots. Their energy 
difference is simply the separation $\d$ between the nearest charge degeneracy 
points. Since this is now a
non-interacting system, the Green's functions $G_{12}$ of the current 
formula \ref{J2site} can be calculated exactly. It is also possible to 
include the effect of non-negligible dot-reservoir and inter-dot coupling.
This yields 
\begin{equation}
J = {e \over h} \int d\w {\g_L \g_R V_M^2 (f_L-f_R) \over
\left[ \w^2-(\d^2+4 V_M^2+\g_L \g_R)/4 \right]^2 + {1 \over 4}
\left[ \w (\g_L+\g_R) + {\d \over 2}(\g_L-\g_R) \right]^2}
\label{2levapp}
\end{equation}

\noindent
First, consider the peak characteristics of the Ohmic conductance. 
This is obtained simply by differentiating with respect to 
$\mu_L$ and setting $\mu_L=\mu_R=\mu$. Thus, for $\g_L=\g_R=\g$, one obtains
\begin{equation}
G = {e^2 \over h} \int d\w 
{1 \over 4 k_BT \cosh^2 \left({\w-\mu \over 2 k_BT} \right)}
{16 \g^2 V_M^2 \over \left[ 4 \w^2 - \d^2-4 V_M^2+\g^2 \right]^2 
+ 4 \g^2 (\d^2 + 4 V_M^2)}
\end{equation}

\noindent
This shows that the conductance is determined by the product of a 
thermal broadening factor and a term which is a Lorentzian in $\w^2$ of 
width $\g \lb$ where $\lb = \sqrt{\d^2+4V_M^2}$. Depending on the relative
width of the two terms, two limiting cases will be of relevance. 

In the limiting case of small reservoir-dot coupling 
$k_BT \gg \sqrt{\lb \g}$ the conductance can be rewritten 
as 
\begin{equation}
G = {e^2 \over \hbar} 
  {\g V_M^2 \over 2 k_BT (\d^2+4 V_M^2+\g^2)} 
  \left\{ 
    \begin{array}{ll} 
       \textstyle 
       \cosh^{-2} \left({\mu \over 2 k_BT}\right) &  \mbox{if $\g \geq \lb$} \\  
       \textstyle \sum_{\pm} 
       {1 \over 2} \cosh^{-2} \left({\mu \pm{1\over2}\sqrt{\d^2+4 V_M^2-\g^2} 
       \over 2 k_BT}\right)  & \mbox{if $\g < \lb$}
    \end{array}
  \right.
\end{equation}
which is clearly consistent with expression \ref{cohinfexp} for negligible 
reservoir-dot and inter-dot coupling. This shows that, as a result of the 
level repulsion and the broadening, the 
conductance does in fact not diverge at $\d=0$ in this limit. Moreover, 
the peak height is inversely proportional to the temperature, as was the
case for the Ohmic conductance through a single dot.

The question whether the conductance diverges at all will now be considered
by taking the opposite limit of vanishing temperature $k_BT \ll \sqrt{\lb \g}$. 
It appears that the conductance either has a single peak or a split peak.
\begin{equation}
G_{\rm max} = {e^2 \over h} 
  \left\{ 
   \begin{array}{lll} 
     \displaystyle {16 \g^2 V_M^2 \over (\d^2 + 4 V_M^2 + \g^2)^2}
        & \mbox{at $\mu=0$} & \mbox{\quad if $\g \ge \lb$} \\

     \displaystyle {4 V_M^2 \over \d^2 + 4 V_M^2}
        & \mbox{at $\mu=\pm {1\over 2} \sqrt{\d^2+4 V_M^2-\g^2}$} &
          \mbox{\quad if $\g < \lb$} 

   \end{array}
  \right.
\end{equation}
 
\noindent
This shows that the conductance will not diverge but has a maximum value
imposed by the conductance quantum $e^2/h$, which can only be reached
when $\d=0, \g_L=\g_R=\g$ and $V_M \ge \g/2$. For asymmetric tunnelling 
barriers to the reservoirs, the conductance is maximised at 
$V_M = {1\over 2}\sqrt{\g_L^2 + \g_R^2}$ and gives a maximum conductance of 
$(e^2/h) 8 \g_L \g_R (\g_L^2+\g_R^2)/(\g_L+\g_R)^4 $.

Secondly, an expression will be derived describing the current peaks which 
occur at finite bias when the gate voltage of one of the dots is raised. 
As before, only a single level per dot will be taken into account. Assuming 
that the energy window $\mu_L-\mu_R$ is sufficiently large, the integral in 
the current equation \ref{2levapp} can be solved analytically by finding 
the residues of the poles of the integrand in the Argand plane. 
This yields
\begin{equation}
J_{\rm peak} = {e \over \hbar} 
{V_M^2 (\g_L+\g_R) \over \d^2 + (\g_L+\g_R)^2/4 + V_M^2 (\g_L+\g_R)^2/\g_L\g_R}
\end{equation}
The current peak has a Lorentzian line shape with a width that is at least
as large as the combined width of the single particle levels in the two dots.
This is consistent with the experimentally observed line shape \cite{Vaart}. 
In the limit of large coupling the same maximum current results as for the 
single dot case. 
}

\section{Summary}
{
\noindent
In conclusion, a coherent system of one or more weakly coupled quantum dots 
with Coulomb 
interaction can be well described using Green's functions of a probabilistic 
nature. 
In a double dot, the modes of the individual dots are coupled together. 
The resulting level repulsion can lead to a complex form of the Ohmic
conductance and the current. Regions of negative differential conductance 
are likely to occur. 
The Ohmic conductance through two dots with multiple levels is 
dominated by the matching energy levels. 

The maximum conductance through a double dot $e^2/h$ is reached at low 
temperatures in the 
symmetric case of aligned levels $\d=0$, identical coupling to the reservoirs
$\g_L=\g_R=\g$ and $V_M \ge \g/2$. 

The peaks that occur in the current as the gate potential of one of the 
dots is varied while keeping the voltage bias across the dots fixed have 
a Lorentzian line shape with a width at least as large as the combined 
broadening of the individual dot levels.

 \begin{figure}
     \center{\hspace{0cm}
         \epsfxsize = 5in
         \epsffile{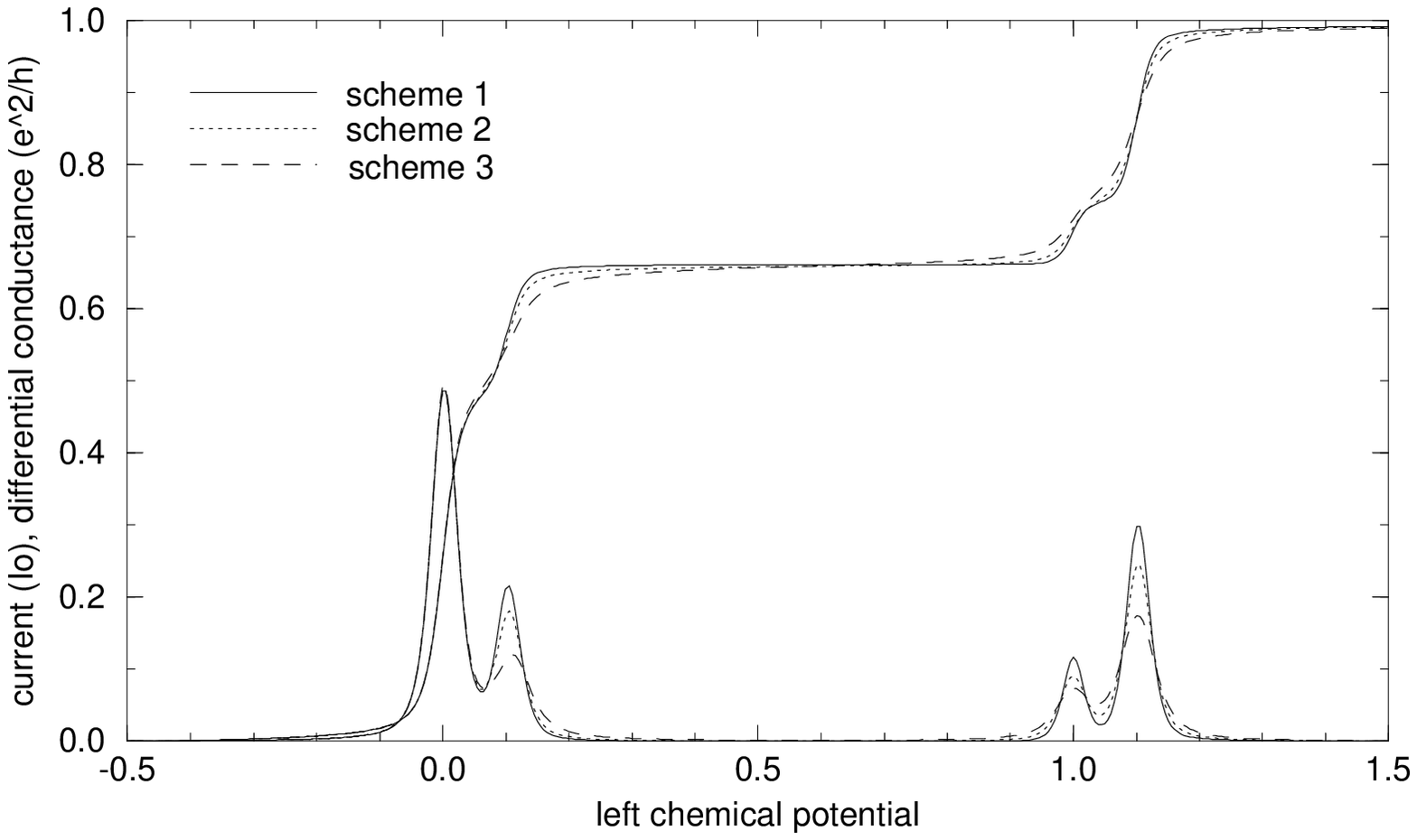}}
    \caption{Current and associated differential conductance through a dot 
       with a spin-split level for the various approximation schemes 
       ($\mu_R = -0.5 U$, $k_BT=\g_L=\g_R=0.01 U$, $\e_{\a}=0.0 U$,
      $\e_{\bar \a} = 0.1 U$, $I_0 = 2 e^2 \g_L \g_R/\hbar(\g_L+\g_R)$)}
    \label{fig:currsch}
\end{figure} 

\begin{figure}
     \center{\hspace{0cm}
         \epsfxsize = 5in
         \epsffile{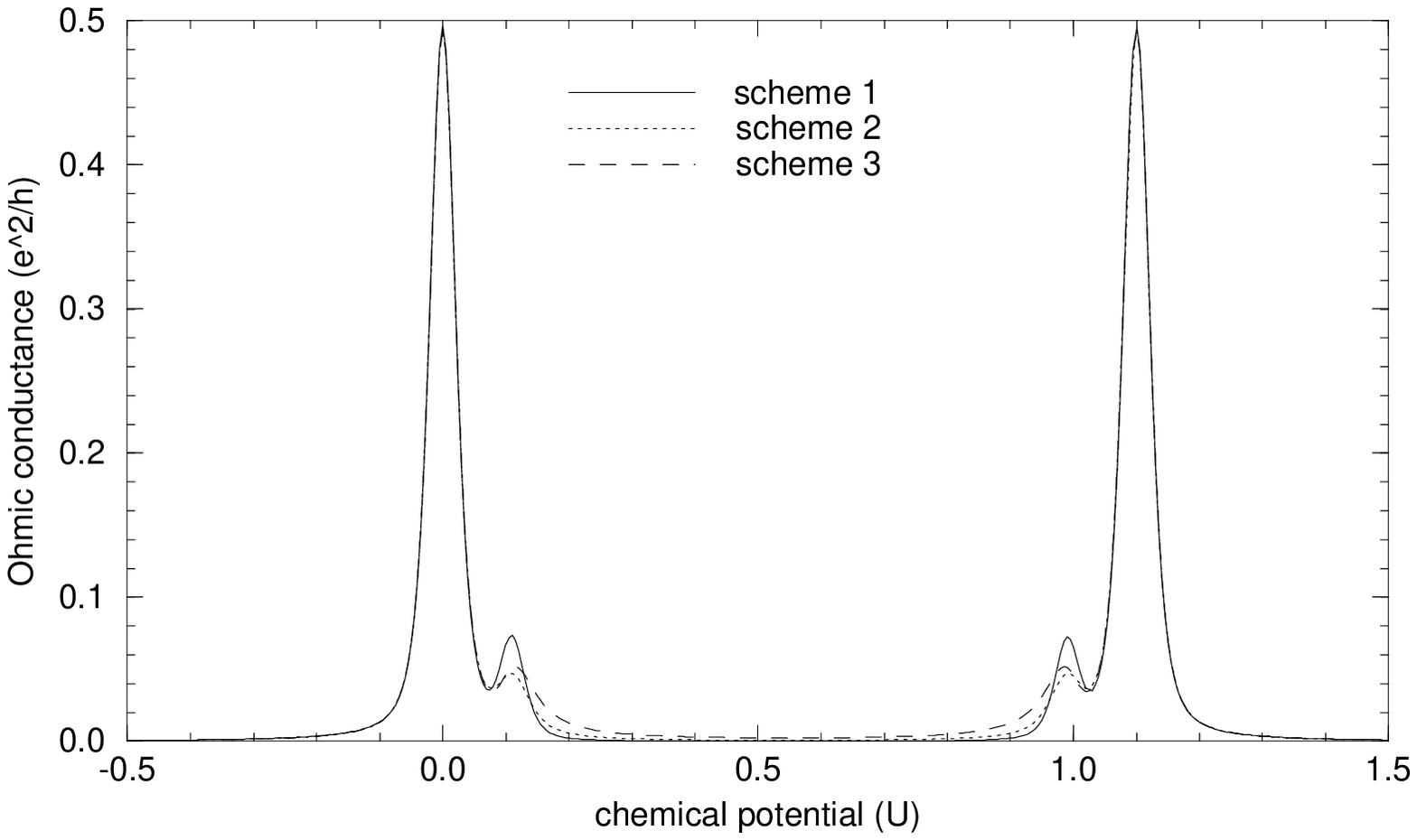}}
    \caption
      {Ohmic conductance through a dot with a spin-split level for the 
      various approximation schemes ($k_BT=\g_L=\g_R=0.01 U$, $\e_{\a}=0.0 U$,
      $\e_{\bar \a} = 0.1 U$).}
    \label{fig:condsch}
\end{figure}  

\begin{figure}
     \center{\hspace{0cm}
         \epsfxsize = 5in
         \epsffile{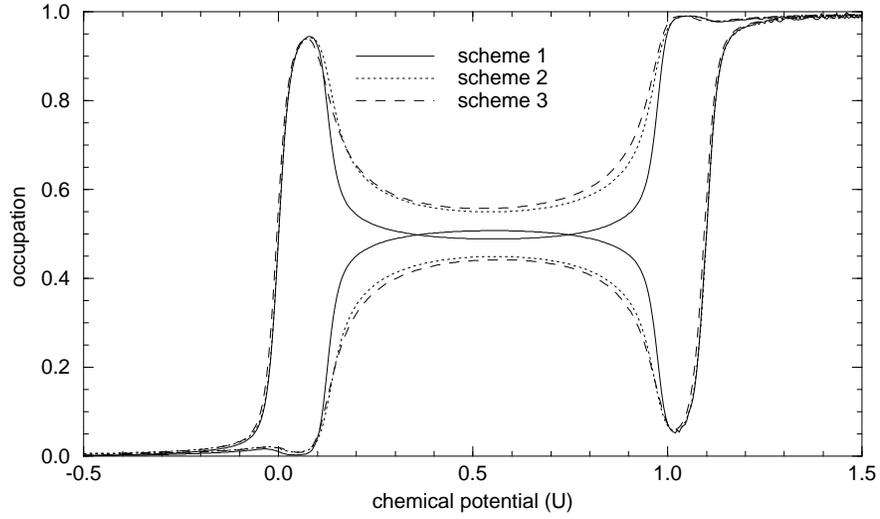}}
    \caption
     {Occupation of a dot with a spin-split level for the 
      various approximation schemes ($k_BT=\g_L=\g_R=0.01 U$, $\e_{\a}=0.0 U$,
      $\e_{\bar \a} = 0.1 U$). The two sets of curves refer to the occupation
      of the $\a$ and $\bar\a$ states.}
    \label{fig:occsch}
\end{figure}  

\begin{figure}
     \center{\hspace{0cm}
         \epsfxsize = 5.8in
         \epsffile{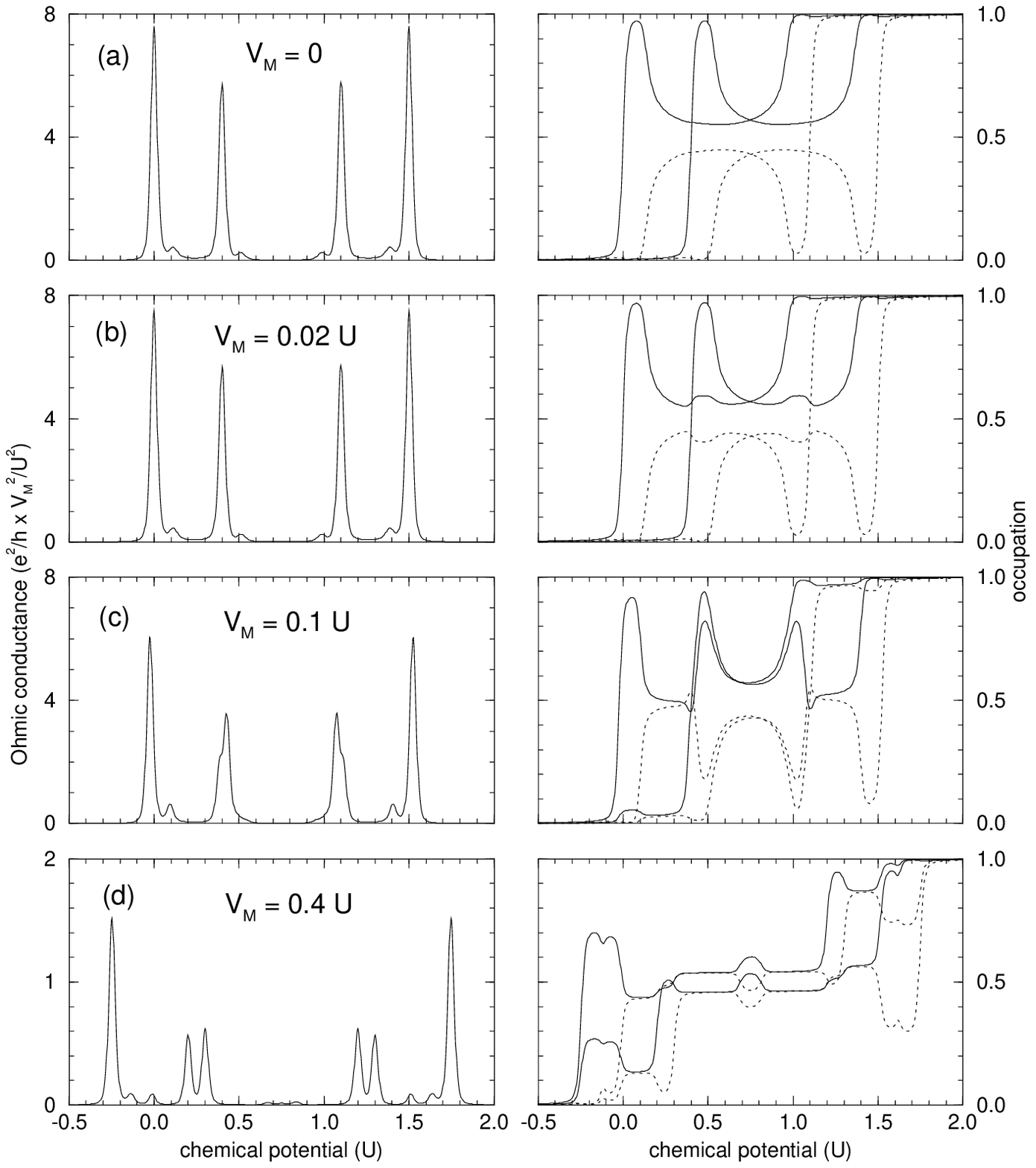}}
    \caption
    {Ohmic conductance through two dots each with a level for both spins,
    using various coupling constants $V_M$
    ($U_1=U_2=U$, $k_BT=\g_L=\g_R= 0.01 U$, $\e_{1\a} = 0.0 U$, 
     $\e_{1\bar\a} = 0.1 U$, $\e_{2\a} = 0.4 U$, $\e_{2\bar\a} = 0.5 U$).
    In the occupation plots the solid lines correspond to the states 
    of spin $\a$, the dotted lines to the states of spin $\bar\a$.}
    \label{fig:cond2s}
\end{figure}
 
\begin{figure}
     \center{\hspace{0cm}
         \epsfxsize = 5.8in
         \epsffile{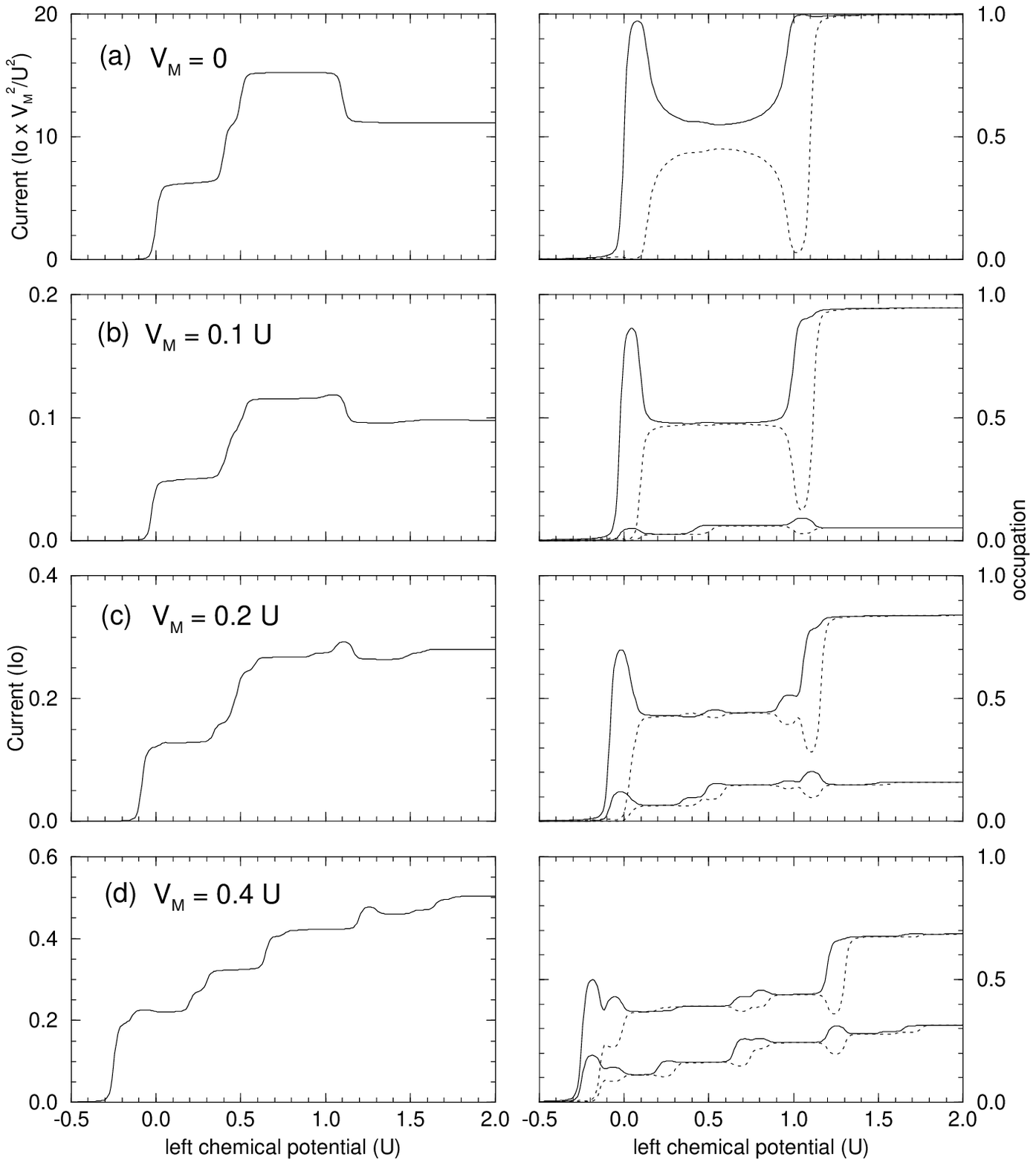}}
    \caption
    {Current through two dots each with a level for both spins,
    using various coupling constants $V_M$
    ($\mu_R = -2.0 U$, $U_1=U_2=U$, $k_BT=\g_L=\g_R= 0.01 U$, $\e_{1\a} = 0.0 U$, 
     $\e_{1\bar\a} = 0.1 U$, $\e_{2\a} = 0.4 U$, $\e_{2\bar\a} = 0.5 U$).
    In the occupation plots the solid lines correspond to the states 
    of spin $\a$, the dotted lines to the states of spin $\bar\a$.}
    \label{fig:curr2s}
\end{figure}

\begin{figure}
     \center{\hspace{0cm}
         \epsfxsize = 5.8in
         \epsffile{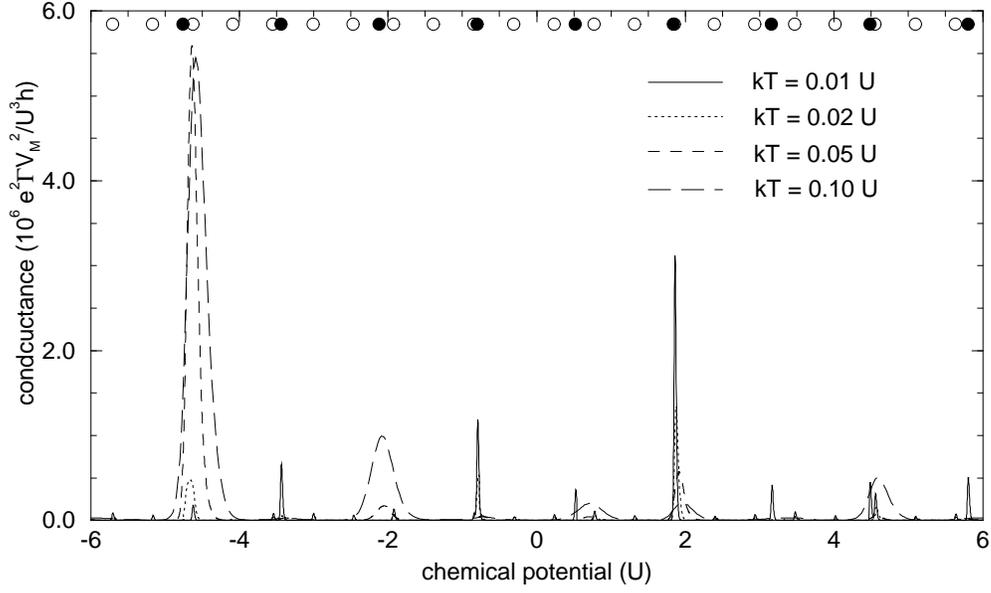}}
    \caption{Ohmic conductance through two dots with multiple levels
             in the limit of small coupling $\g,V_M$
             ($U_1=0.41 U$, $U_2 = U$, $\d_{e_1} = 0.13 U$, $\d_{e_2} = 0.32 U$, 
             $\g_L=\g_R=\g$). The empty and filled circles indicate the 
             positions at which the average occupation increases by one 
             for dot 1 and 2 respectively.}
    \label{fig:cohinf}
\end{figure}

\end{document}